\RequirePackage{fix-cm}
\documentclass{iopjournal}

\usepackage{amsmath,amssymb}
\usepackage{bm}
\usepackage{braket}
\usepackage{url}
\usepackage{cite}
 
\begin{document}

\title{Digital-Analog Quantum Neural Network with Long-Range Interactions}

\author{Zongyuan Ge$^{1,2,*}$, Daili Li$^{1,3,\dagger}$ and Zewen Zhang$^{1}$}

\affil{$^1$ Shanghai Qi Zhi Institute, Shanghai 200232, China}

\affil{$^2$ Department of Physics, National University of Singapore, Singapore 117551, Singapore}

\affil{$^3$ School of Physics and Key Laboratory of Quantum State Construction and Manipulation (Ministry of Education), Renmin University of China, Beijing 100872, China}

\email{$^*$gezongyuan@u.nus.edu, ~$^\dagger$lidaili@ruc.edu.cn}

\begin{abstract}

Protocols combining single-qubit digital operations with analog quantum evolutions bypass large numbers of two-qubit entangling gates, offering a promising approach to variational algorithms and hybrid quantum simulation. Focusing on the digital-analog quantum neural network (DA-QNN) as a representative framework, this work demonstrates that incorporating long-range interactions enhances the efficiency of such hybrid algorithms. Within the analog blocks, an Ising model with power-law interactions is employed to generate entanglement, and this protocol is benchmarked across supervised and unsupervised learning tasks. Beyond discussing the effects of the transverse field and evolution time, the primary analysis focuses on how the interaction range governs the expressibility and trainability of the ansatz. For supervised image classification and many-body phase detection, long-range ansatzes match the performance of their short-range counterparts with fewer DA-QNN layers or shorter evolution times, owing to the rapid information propagation across the system enabled by nonlocal couplings. This resource reduction alleviates the classical optimization overhead and eases experimental implementation. In unsupervised phase detection, although the long-range ansatz does not outperform its short-range counterpart in every regime, the two ansatzes provide complementary signatures that jointly reconstruct the complete phase diagram. These findings establish long-range interacting platforms as a compelling resource for advancing practical, large-scale digital-analog quantum computing.
\end{abstract}

\section{Introduction}
\label{sec:intro}

Variational quantum circuits (VQCs) have demonstrated broad potential across diverse areas, including combinatorial optimization~\cite{farhi2014quantum,harrigan2021nonplanar,ebadi2022maximum,diezvalle2023qaoa,zhu2023multiround,apte2026integer}, quantum
chemistry~\cite{whitfield2011simulation,peruzzo2014variational,mcclean2016theory,kandala2017hardware,grimsley2019adaptive,zhuang2024hardware}, and machine learning~\cite{lu2024digital,liu2025hybrid,khatun2026hardware,boneberg2026getting,nakano2026neural}. Conventional digital ansatzes, however, require numerous entangling gates to build sufficient
entanglement~\cite{shende2004minimal,zhang2025grover}, and the resulting circuit depth is prohibitive on near-term hardware. For instance, two-qubit gate durations on trapped-ion platforms range from a few tens to hundreds of microseconds, reflecting a speed--fidelity trade-off~\cite{ballance2016high,bruzewicz2019trapped,ransford2026trapped}, and parallelizing these operations remains challenging~\cite{grzesiak2020efficient,zhu2023pairwise,diaz2026arbitrary}. For neutral-atom platforms, two-qubit gates remain the dominant error source even at state-of-the-art fidelities~\cite{evered2026high,senoo2026high}, which restricts practical VQC depths and imposes substantial error-mitigation or error-correction overheads. Meanwhile, on superconducting-circuit platforms, compiling nonlocal entangling gates requires extensive SWAP operations~\cite{arute2019quantum,ge2026enhanced}  due to rigid connectivity constraints. Therefore, a fast yet controllable mechanism for generating entanglement is essential for realizing practical VQCs.

Hardware-efficient algorithms~\cite{kandala2017hardware}  mitigate this gate-count overhead by exploiting native hardware properties, including both the connectivity and intrinsic dynamics. Recently, digital-analog (DA) ansatzes~\cite{arrazola2016digital,lamata2018digital,rajabi2019dynamical,kokail2019self,parrarodriguez2020digital,gonzalezraya2021digital,michel2023blueprint,richerme2025multimode,deshpande2026analog,lamata2026digital,schneider2026universal} have been proposed to replace resource-intensive entangling blocks with the unitary evolutions of a native many-body Hamiltonian. These protocols have been applied to both quantum simulation, where a small set of digital controls extends a native many-body Hamiltonian beyond its vanilla dynamics~\cite{lamata2018digital,katz2025hybrid,kiefer2026digital}, and near-term algorithms, where analog entangling blocks reduce the cost of compiling digital gates~\cite{martinez2016compiling,maslov2018use,martin2020digital,zhuang2024hardware,lu2024digital,kumar2025digital}. Such DA implementations span all major platforms, including Rydberg atom arrays~\cite{michel2023blueprint,lu2024digital}, trapped ions~\cite{rabinovich2022ion,zhuang2024hardware,kumar2025digital}, and superconducting circuits~\cite{lamata2018digital,gonzalezraya2021digital,andersen2025thermalization}.
Across these applications, the ansatz expressibility  requires  analog blocks to spread  information efficiently, whereas practical  trainability demands that the system be robust~\cite{mcclean2018barren,park2024hamiltonian,losl2026variational}. In this context, long-range interacting models, whose couplings decay algebraically~\cite{defenu2023long}, exhibit both rapid information scrambling~\cite{hauke2013spread,schachenmayer2013entanglement,pappalardi2018scrambling,PhysRevB.101.020416,joshi2022observing} and slow thermalization~\cite{mukamel2005breaking,gong2013prethermalization,mori2019prethermalization}, making them well suited for DA protocols. The long-range  transverse-field Ising
model (LR-TFIM), which is native to trapped ions~\cite{kim2010quantum,islam2013emergence,jurcevic2017direct,zhang2017observation,li2023probing}, serves as a competitive candidate for the analog blocks in a digital-analog quantum neural network (DA-QNN). Other platforms, such as  dipolar Rydberg arrays~\cite{deleseleuc2017optical,orioli2018relaxation,signoles2021glassy,chen2023continuous} and ultracold polar molecules~\cite{ni2010dipolar,yan2013observation,christakis2023probing,gregory2024second}, also provide long-range interacting models.
Furthermore, the tunable interaction range across these platforms~\cite{britton2012engineered,tabares2026programming} provides high flexibility for practical DA-QNN design.

Variational protocols incorporating  analog evolutions with long-range interactions have been employed for fast state preparation~\cite{ho2019ultrafast}, combinatorial
optimization~\cite{pagano2020quantum}, and variational quantum eigensolvers (VQEs)~\cite{zhuang2024hardware}. In these
studies, however, the interaction range is either fixed intrinsically by the platforms (\textit{e.g.},
a power-law exponent $\alpha=6.0$ by the van der Waals tail on Rydberg
arrays~\cite{lu2024digital}) or restricted to a single representative value~\cite{zhuang2024hardware,martinezpena2024quantum}. Consequently, how the interaction range itself governs performance in quantum
learning tasks has  not been systematically investigated.

This work combines  fast state preparation via nonlocal interactions with hardware-efficient entanglement layers in DA ansatzes, and especially focuses on the effect of the interaction range in DA-QNNs.
Rather than fixing $\alpha$~\cite{zhuang2024hardware}, this work treats it as a tunable hyperparameter for  ansatz construction to systematically explore its impact. Across representative learning tasks, $\alpha$ is swept from $0.2$ to $6.0$, spanning the interaction regimes natively realized in leading quantum hardware, with the nearest-neighbor limit included for a baseline comparison.
One of the major findings of this work is that the ansatz with long-range interactions requires significantly fewer resources,  either fewer DA-QNN layers or shorter physical execution times, than its short-range counterparts, which highlights its advantage over previous DA-QNN ansatzes. 
Because this performance gap widens with system size, long-range interactions hold great potential for future large-scale implementations.

Specifically, the DA-QNN with long-range interactions is applied to a broad range of supervised and unsupervised  tasks, demonstrating consistently stable performance. Besides the binary classification of handwritten digits from the MNIST dataset, where the long-range interactions exhibit a clear advantage,  the same ansatzes are also applied to supervised phase classification in a cluster symmetry-protected topological (SPT) model and unsupervised phase characterization in an XXZ spin chain. In these many-body tasks, although the long-range ansatzes do not outperform their short-range counterparts in every setting, they provide crucial complementary signatures for sorting quantum phases, underscoring their utility across diverse learning regimes.
While this work primarily focuses on the interaction range, our benchmarking also extends to other key parameters such as the transverse field and circuit depth, as well as the data-encoding strategy. 
This tunability closely aligns with the complex thermalization in the dynamics of long-range interacting systems and facilitates the optimization of physical execution times. 

The paper is structured as follows: Section~\ref{sec:ansatz} introduces the DA-QNN ansatz with long-range interactions. Section~\ref{sec:main} presents the results of this ansatz on binary MNIST classification, scanning the interaction range and other settings. Section~\ref{sec:phase-detection} further applies the ansatz to supervised cluster-SPT classification and unsupervised XXZ characterization and benchmarks the role of long-range interactions in a broader scope.
Section~\ref{sec:conclusion} concludes the work.

\section{Ansatz}
\label{sec:ansatz}

\subsection{LR-TFIM Hamiltonian}
\label{sec:hamiltonian}

The analog block of the circuits considered in this work is an analog evolution under the LR-TFIM on a chain of $N$ qubits
\begin{equation}
H(\alpha,h) = -\,\frac{J_0}{\mathcal{N}_\alpha}\sum_{i<j}
 \frac{\sigma^z_i\sigma^z_j}{|i-j|^\alpha}
 -h\sum_i \sigma^x_i,
\label{eq:H}
\end{equation}
where $\sigma^z_i$ and $\sigma^x_i$ denote the Pauli-$z$ and $x$ matrices for the $i^\text{th}$ qubit, respectively, $\alpha$ is the power-law exponent, and $h$ represents the transverse field. The interaction energy scale is set to $J_0=1$. A Kac normalization factor
$\mathcal{N}_\alpha =
\frac{1}{N-1}\sum_{i<j}|i-j|^{-\alpha}$~\cite{homrighausen2017anomalous} and open boundary conditions are applied to reflect realistic hardware settings.
In what follows, we use the evolution time $\tau$ and fix it for all analog blocks within a given ansatz.

\subsection{Quantum neural network protocol}
\label{sec:protocol}
The ansatz alternates between two types of layers~\cite{parrarodriguez2020digital,lu2024digital}:
a digital layer of parameterized single-qubit rotations
$R(\boldsymbol{\theta})$ with Euler angles $\boldsymbol{\theta}$, and an analog
block with global evolution $e^{-iH\tau}$ generated by the LR-TFIM of
Eq.~\eqref{eq:H}. Instead of realizing
an arbitrary entangling unitary, the analog block has only three physical parameters: the power-law exponent $\alpha$, the transverse field $h$, and the
evolution time $\tau$, which together control the information propagation.
Accordingly, a depth-$\ell$ circuit interleaves $\ell$ analog blocks with
$\ell+1$ digital rotation layers,
\begin{equation}
U(\boldsymbol{\theta},\tau)
= R_{\ell}(\boldsymbol{\theta}_{\ell})\,e^{-iH\tau}\,
 R_{\ell-1}(\boldsymbol{\theta}_{\ell-1})\cdots e^{-iH\tau}\,
 R_{0}(\boldsymbol{\theta}_{0}),
\label{eq:ansatz}
\end{equation}
where $R_j(\boldsymbol{\theta}_j)$, $j=0,\dots,\ell$, realizes an arbitrary single-qubit rotation on
the Bloch sphere of each qubit. In total, the circuit carries $3N(\ell+1)$ real variational parameters.

\begin{figure}[!htbp]
\centering
\includegraphics[width=\textwidth]{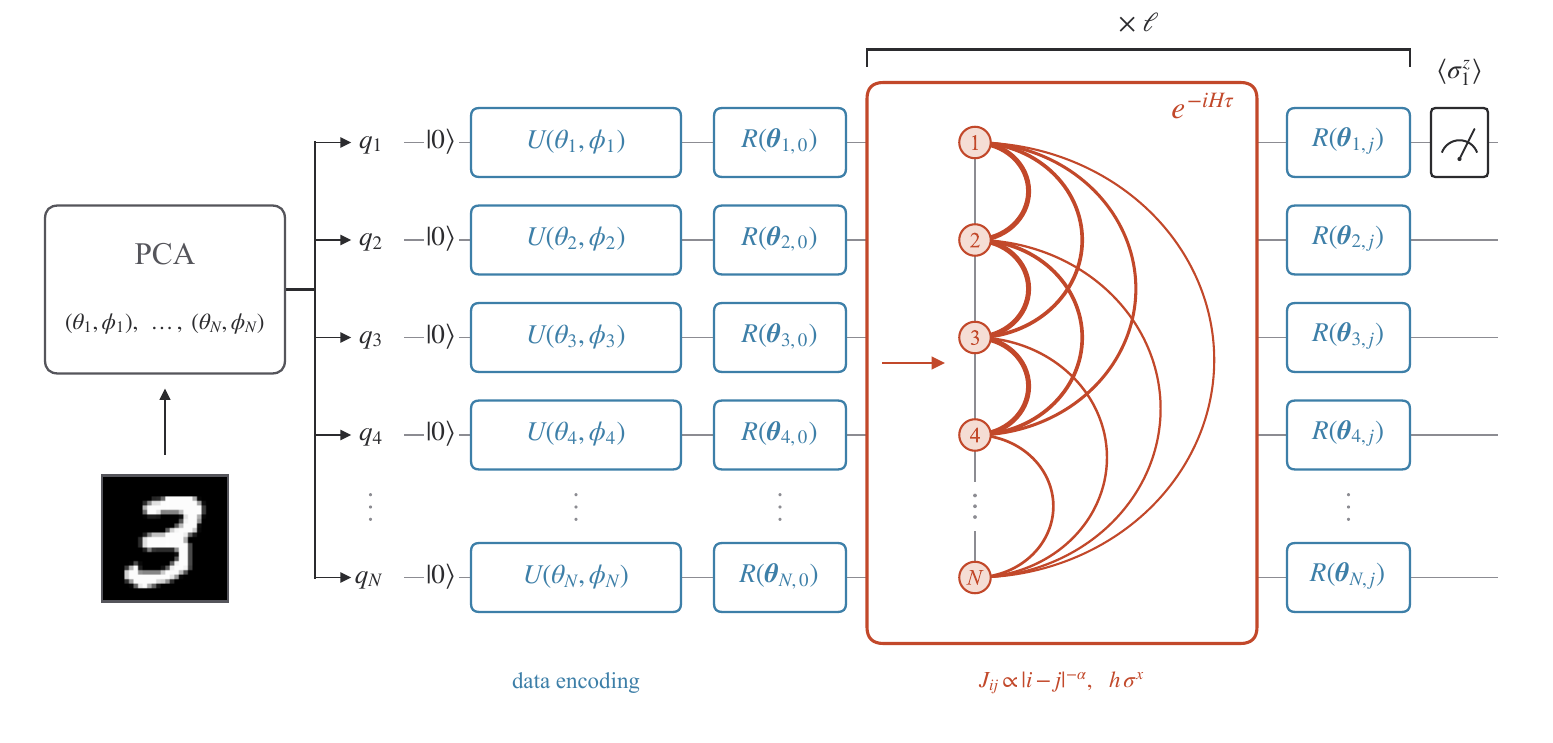}
\caption{DA-QNN pipeline. An MNIST image is decomposed via PCA into  $2N$ components
$\{(\theta_q,\phi_q)\}$ and  encoded  on qubits
through Bloch-sphere rotations $U(\theta_q,\phi_q)$.  The encoded state is then processed by alternating digital and analog operations. Blue boxes denote the encoding rotations $U(\theta_q,\phi_q)$ and the
digital single-qubit rotations $R(\boldsymbol{\theta}_{q,j})$ of
Eq.~\eqref{eq:ansatz}; the red box denotes an analog 
evolution under the LR-TFIM parameterized by time $\tau$, in which the arcs indicate the power-law couplings and the horizontal arrow indicates the transverse field $h$. One of the qubits is measured at the end for classification.}
\label{fig:architecture}
\end{figure}

The typical workflow of our DA-QNNs is demonstrated in Figure~\ref{fig:architecture} and exemplified with a binary classification task. Each input sample is represented by $2N$ real numbers and encoded on the Bloch spheres of the qubits~\cite{larose2020robust}. For example,
MNIST $28\!\times\!28$ images are reduced by principal component analysis (PCA) to $2N$ components and
encoded into a product state $\bigotimes_{q=1}^{N}
\bigl[\cos(\theta_q/2)\ket{0}+e^{i\phi_q}\sin(\theta_q/2)\ket{1}\bigr]$,
with the polar angles $\{\theta_q\}$ and azimuthal angles $\{\phi_q\}$ of
the $N$ qubits carrying the $2N$ PCA components. We order the principal components by
importance and assign the first $N$ components to
the polar angles $\{\theta_q\}$ and the remaining $N$ components to the
azimuthal angles $\{\phi_q\}$, starting from the central qubit and
moving outward. Other encoding schemes are also investigated in Section~\ref{sec:encoding}.
The readout qubit $q_1$ is measured in the computational basis, and the image is
assigned to one of the two digits according to whether 
$p_0=\tfrac{1}{2}(1+\langle \sigma^z_1\rangle)$ exceeds $1/2$.
The test accuracy 
is the fraction of test images assigned to the correct labels.

Training uses AdaGrad~\cite{duchi2011adaptive} with a learning rate of $0.1$ to minimize the binary cross-entropy
$-\bigl[(1-y)\log p_0+y\log(1-p_0)\bigr]$, where $y\!\in\!\{0,1\}$ denotes the true class of the image. Analytic gradients of the loss with respect to the variational parameters are obtained by reverse-mode automatic differentiation through an exact simulation of the quantum state~\cite{luo2020yao}. Each training process takes $70$ iterations of parameter updates, and each iteration randomly selects $100$ images out of a balanced training set of $1000$ images ($500$ per class). The trained circuit is then evaluated on a disjoint test set of $200$ images ($100$ per class). Optimization results are averaged over $10$ random seeds per trial in most of this work, while in Sections~\ref{sec:depth-time} and \ref{sec:alphaN} and the corresponding appendices, we use $20$ seeds to reduce the statistical uncertainty. Unless stated otherwise, error bars are $95\%$ $t$-based confidence intervals of the mean over the random seeds.

\section{Binary classification task}
\label{sec:main}

To identify the role of the interaction range in DA-QNNs, we use the binary MNIST classification as the first testbed, focusing primarily on the digit pair 3 versus 8 (written 3v8 hereafter). This pair is a moderately challenging task among the classifications we examined (see Appendix~\ref{app:task-diff}), thus providing a discriminating benchmark for ansatzes with various interaction ranges. A broad parameter scan in $(\tau,\ell)$ is performed in Section~\ref{sec:fixed-tl} to ensure that the parameters besides $\alpha$ are in an appropriate range. Section~\ref{sec:depth-time} shows that the ansatz with long-range interactions reaches a target accuracy with shorter analog evolution time or lower circuit depth than its short-range counterpart. Section~\ref{sec:alphaN} reports that the long-range advantage grows with system size $N$. Section~\ref{sec:encoding} discusses the encoding strategy.

\subsection{Preliminary scan of circuit depth and analog block duration}
\label{sec:fixed-tl}

In the DA-QNN setup, the duration $\tau$ of the analog block plays an important role in  performance.
It should be long enough to allow information to propagate across the constituent qubits, yet short enough to prevent full thermalization and to ensure experimental viability. To determine an appropriate setting of $\tau$ for the ansatzes, a scan is performed over
the $(\tau,\ell)$ grid at $N\!=\!8$ and $\alpha\!=\!1.0$, sweeping $\tau$ from $0.2\pi$ to $1.6\pi$ in steps of $0.2\pi$ and $\ell\!\in\!\{1,2,3,4,6,8,10\}$ for both $h\!=\!1.0$ and $h\!=\!3.0$.

\begin{figure}[!htbp]
\centering
\includegraphics[width=0.65\textwidth]{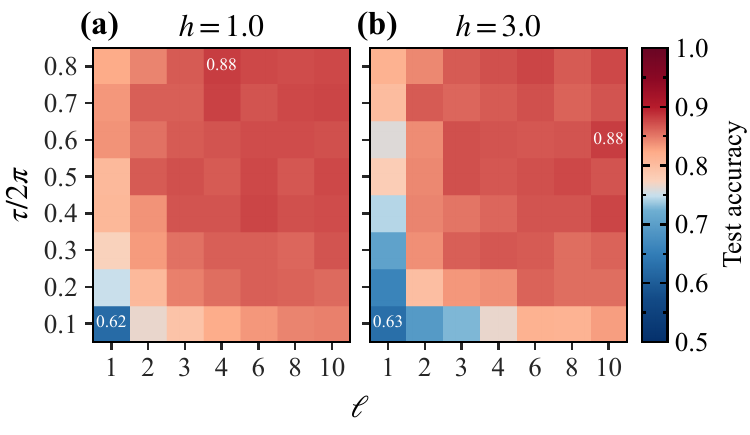}
\caption{Binary classification accuracy of the digit pair 3v8 across $(\tau,\ell)$ values for
(a)~$h\!=\!1.0$ and (b)~$h\!=\!3.0$ at $N\!=\!8$ and $\alpha\!=\!1.0$. Results are averaged over random seeds. The best and worst cells of each panel are annotated.}
\label{fig:tl-heatmap}
\end{figure}

As shown in Figure~\ref{fig:tl-heatmap}, for shallow circuits with small $\ell$, the test accuracy improves as the evolution time of the analog blocks becomes longer. This is because information needs time to propagate across the system, even though long-range interacting systems do not have a linear lightcone to constrain such propagation. For circuits with larger $\ell$, the advantage of a long evolution time is only marginal, as information propagation accumulates over layers and becomes adequate for attaining high expressibility. 
The trends are consistent for both $h$ values. As detailed in Appendix~\ref{app:therm} and discussed extensively in the literature~\cite{mukamel2005breaking,gong2013prethermalization,mori2019prethermalization}, the thermalization rate is determined by $\alpha$ and $h$. 
To ensure the system at $h=1.0$ is not thermalized within one or a few QNN layers, for consistency we adopt $\tau=0.4\pi$ as the baseline for the MNIST tasks and the extensions below, unless specified otherwise.

At small $\tau$, as is expected for most variational ansatzes, adding layers continues to improve the accuracy until it saturates. The number of layers needed to reach saturation decreases as $\tau$ grows, suggesting that this saturation arises from the dynamical properties of the analog part rather than from barren plateaus. Since the relaxation time of long-range interacting systems grows with system size~\cite{kastner2011diverging,zhang2026interplay}, the circuit is expected to remain unsaturated at larger $\ell\tau$ values as $N$ increases.
Based on the observations presented in this section and the performance scan over $(\alpha,h)$ in Appendix~\ref{app:h-alpha}, this work focuses on the results for $h=1.0$ and $\tau=0.4\pi$, but the conclusions  here are generally applicable to other transverse field strengths.

\subsection{Minimal resources for each interaction range}
\label{sec:depth-time}

\begin{figure}[!htbp]
\centering
\includegraphics[width=0.65\textwidth]{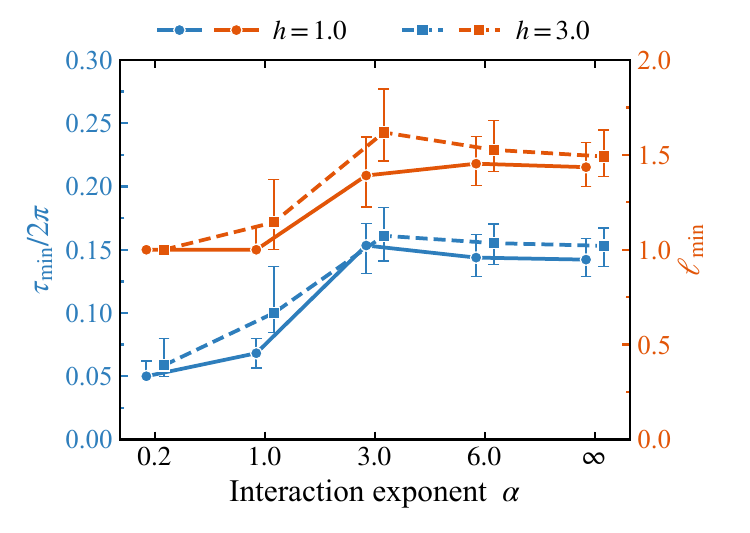}
\caption{Minimum resources to reach $\mathrm{acc}^\star\!=\!0.70$ versus $\alpha$ on the pair 3v8 with $N\!=\!8$. Blue: the shortest layerwise evolution time
$\tau_{\min}$. Orange: the lowest depth $\ell_{\min}$. Solid
circles: $h\!=\!1.0$, dashed squares: $h\!=\!3.0$. 
Error bars are $95\%$ bootstrap confidence intervals over random seeds.
Markers have been shifted slightly along the horizontal axis for better visibility.
}
\label{fig:resource}
\end{figure}

Incorporating long-range interactions into DA-QNNs substantially reduces computational expenditure.
To evaluate the resources needed for DA-QNNs to reach a target accuracy $\mathrm{acc}^\star$, two comparisons are made in this subsection: one scans the minimal time $\tau_{\min}$ at $\ell\!=\!2$ to reach $\mathrm{acc}^\star$, and the other scans the minimal number of QNN layers $\ell_\text{min}$ at a fixed evolution time $\tau=0.4\pi$.  Without loss of generality, the criterion  for comparison is set to $\mathrm{acc}^\star\!=\!0.70$ here, and $\tau_{\min}$ is located via a grid search with a step of  $0.1\pi$. Figure~\ref{fig:resource} presents the comparisons using the pair 3v8 at $N\!=\!8$, and the detailed curves of accuracy versus time and layers are given in Appendix~\ref{app:resource-curves}. 
In Figure~\ref{fig:resource}, for both transverse fields, the analog block duration required by the DA-QNN with $\alpha=0.2$ is shorter than that of the other ansatzes. In short-range interacting systems, information propagation is linearly bounded, so that the spreading of information across the chain takes a long time. In contrast, long-range interacting systems propagate information faster.
With respect to circuit depth, the ansatz with $\alpha\!=\!0.2$ reaches $\mathrm{acc}^\star$ within a single layer at both fields and the one with $\alpha\!=\!1.0$ needs at most about $1.1$ layers on average, whereas the short-range ansatzes need around $1.5$ layers.

This reduction in resources offers multiple advantages, including lower optimization overhead and enhanced robustness against noise for future implementation on quantum hardware.
On the one hand, fewer layers require fewer parameters to tune, and therefore alleviate the burden imposed on classical optimization routines and provide relief from the notorious barren plateau problem. On the other hand, a shorter physical runtime has two practical advantages: it reduces the time that the analog block is exposed to decoherence on hardware, such as laser fluctuations and nonadiabatic excitations, so that the fidelity of the algorithm is better preserved. Furthermore, given that the runtime of large entangling operations is a bottleneck in state-of-the-art large-scale trapped-ion platforms~\cite{guo2024site,guo2026observation}, the modest entangling time demand of our algorithm benefits its implementation under practical experimental conditions.

\subsection{Size-dependent performance comparison}
\label{sec:alphaN}

\begin{figure}[!htbp]
\centering
\includegraphics[width=0.95\textwidth]{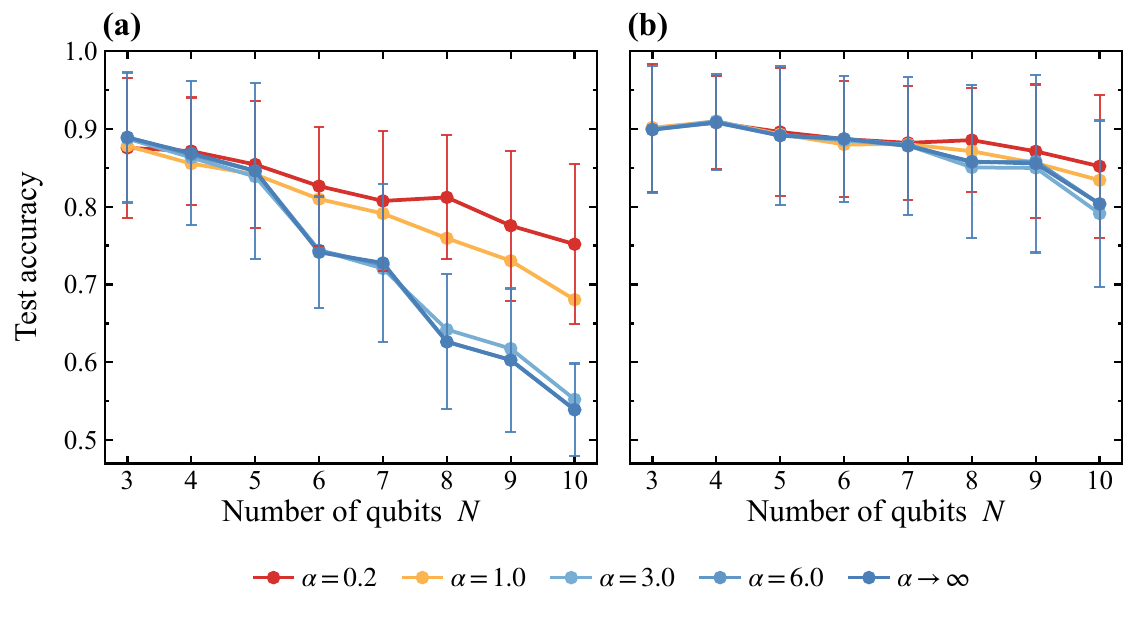}
\caption{Binary classification accuracy with various system sizes $N$ at $h\!=\!1.0$ for (a)~$\ell\!=\!1$ and (b)~$\ell\!=\!2$, averaged over six digit pairs. The curves for $\alpha\!=\!6.0$ and $\alpha\!\to\!\infty$ nearly coincide.
Error bars are one standard deviation across the six pairs and random seeds and shown only for $\alpha\!=\!0.2$ and $\alpha\!\to\!\infty$ for clarity.
}
\label{fig:alphaN-acc}
\end{figure}

The advantage of nonlocal DA-QNN architectures is amplified with increasing system size, as information in long-range interacting systems propagates faster than the linear lightcone that bounds short-range interacting systems~\cite{hauke2013spread,losl2026variational}. 
To verify this intuition, Figure~\ref{fig:alphaN-acc} presents the performance of the DA-QNN with different $\alpha$ for two circuit depths, averaged over six digit pairs (3v8, 0v1, 0v6, 2v7, 5v8, 4v9).
The results for shallow circuits with $\ell\!=\!1$ in Figure~\ref{fig:alphaN-acc}(a) indicate that, at small system sizes ($N\!\leq\!5$), where $\tau$ allows sufficient time for local information to propagate across the whole system, the performance of the DA-QNN exhibits no dependence on $\alpha$. As the system scales up, the ansatz with $\alpha=0.2$ outperforms those with shorter-range interactions, and this advantage grows with $N$. 
For the deeper circuits with $\ell\!=\!2$ in Figure~\ref{fig:alphaN-acc}(b), information propagation is enhanced, which closes the performance gap between different $\alpha$ values for 
$N\!<\!8$. However, as the system size increases, the advantage of long-range interactions reappears at $N\!\geq\!8$.
This performance gap will continue to close at larger 
$N$ as the DA-QNN goes deeper (see Appendix~\ref{app:depth-size}), but the advantage of long-range interactions is expected to persist in future large-scale applications to practical tasks.

The accuracy of all ansatzes decreases as $N$ grows, which is similar to the observation reported in Ref.~\cite{lu2024digital} and does not change the validity of our conclusion regarding the advantage of long-range ansatzes.
This decrease arises because the binary digit classification task is so simple that usually only up to 10 principal components each account for more than $2\%$ of the total variance, and these leading components are sufficient for an ansatz with 3 to 5 qubits to reach high accuracy. Thus, increasing 
$N$ to add more components with small variance helps little in encoding useful information, but instead raises the difficulty of training the variational parameters~\cite{mcclean2018barren,park2024hamiltonian}.

\subsection{Encoding strategies}
\label{sec:encoding}

\begin{figure}[!htbp]
\centering
\includegraphics[width=0.55\textwidth]{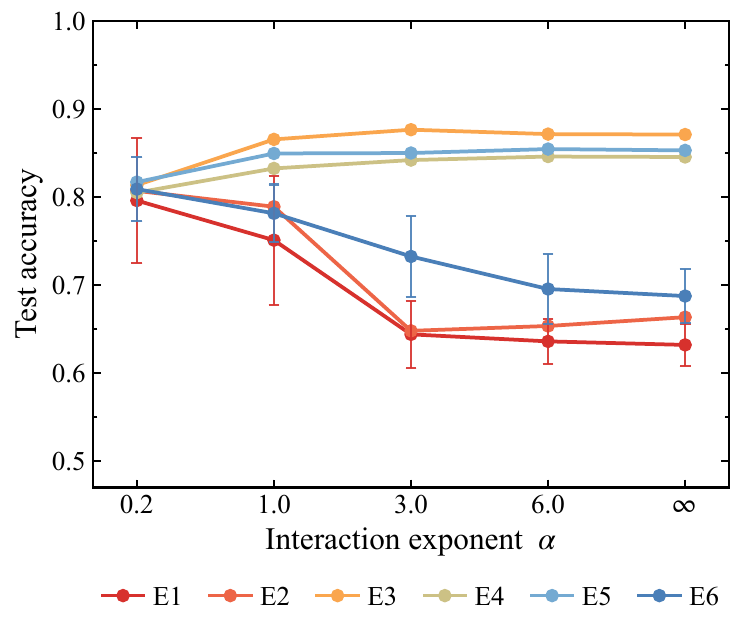}
\caption{Effect of the encoding strategy on binary classification accuracy across $\alpha$ for the pair 3v8 at $N\!=\!8$, $h\!=\!1.0$, and
$\ell\!=\!1$. Each strategy
encodes PCA components in order of importance and assigns a readout qubit: E1 [starting from the central qubit $q_5$ and outward ($q_5\rightarrow q_6\rightarrow q_4\dots q_1$) with readout qubit $q_1$, the scheme used throughout this work], E2 [starting from $q_5$ and outward ($q_5\rightarrow q_4\rightarrow q_6\dots q_1$), $q_1$], E3 [sequential ($q_1\rightarrow q_8$), $q_1$], E4 [($q_5\rightarrow q_6\rightarrow q_4\dots q_1$), $q_5$], E5 [($q_5\rightarrow q_4\rightarrow q_6\dots q_1$), $q_5$], E6 [sequential, $q_5$]. 
Error bars are $95\%$ confidence intervals of the mean over random seeds and shown only for E1 and E6 for clarity.}
\label{fig:encoding-ablation}
\end{figure}

The nonlocal feature inherent in long-range interactions smears out
the differences among individual encoding qubits, which gives the algorithm
robustness against variations in hardware-level design. Most notably, the
DA-QNN with long-range interactions is insensitive to the encoding
strategy, in contrast to its short-range counterparts.
Figure~\ref{fig:encoding-ablation} compares the performance of different encoding strategies for $N=8$, $h=1.0$, and $\ell\!=\!1$. There are two types of encoding order: one that starts from the central qubit and moves outward, and another that proceeds sequentially from the first qubit. Additionally, two readout options are available: either the first qubit or the central qubit.
As described in Section~\ref{sec:protocol}, the encoding in this work begins at the central qubit and progresses outward, with the first qubit used for readout. This protocol follows the ordering in Ref.~\cite{lu2024digital} and is labeled E1 in Figure~\ref{fig:encoding-ablation}.
The comparison among all  combinations of encoding orders and readout choices indicates that the choice of data encoding strategy affects the accuracy
of quantum learning tasks and can itself be 
optimized~\cite{tokuhiro2026discovering}, but long-range interactions close this performance gap  at $\ell=1$. This gap is narrowed but not closed at $h=3.0$, which is not shown here for brevity.

The encoding choice matters only at shallow depths, and as the circuit goes deeper, the differences between long-range and short-range interactions fade away. When $\ell\!=\!1$, for $\alpha\!\geq\!3.0$, the six encodings span a range of about $0.24$ in accuracy. By $\ell\!=\!4$ the
spread of the mean accuracy across the six encodings has already narrowed to
$0.017$ (averaged over $\alpha$), roughly an order of magnitude
smaller than the corresponding value of $0.17$ at $\ell\!=\!1$ (Appendix~\ref{app:depth-encoding}). 
Repeated analog evolutions mix the encoded features across the system, so that
different encoding orders and readout sites no longer affect accuracy.

\section{Quantum phase detection}
\label{sec:phase-detection}

This section compares the performance of the DA-QNN across different interaction ranges in sorting 
quantum many-body ground states.
Common quantum classifiers often use problem-specific architectures inspired by tensor networks or the target Hamiltonian~\cite{cong2019quantum,herrmann2022realizing,catala2026physics,scarato2026hybrid}.
Here, the circuit architecture is kept fixed across all tasks. We apply the DA-QNN to phase detection in two models: the cluster-SPT model and the XXZ model with next-nearest-neighbor interactions.

\subsection{Supervised phase classification on a cluster-SPT chain}
\label{sec:spt}

This subsection benchmarks the DA-QNN across various interaction ranges on supervised phase classification, focusing specifically on a cluster-SPT model with an open-boundary spin-$1/2$ chain~\cite{cong2019quantum,chen2026quantum}
\begin{equation}
    H_{\rm SPT}=-J\sum_{i=1}^{N-2} \sigma^z_i\sigma^x_{i+1}\sigma^z_{i+2}-h_1\sum_{i=1}^{N} \sigma^x_i-h_2\sum_{i=1}^{N-1} \sigma^x_i\sigma^x_{i+1},
\label{eq:cluster-spt}
\end{equation}
where $J$ is the three-body cluster interaction strength, $h_1$ is a uniform transverse
field, and $h_2$ is a nearest-neighbor $XX$ interaction.  The phase diagram of Eq.~\eqref{eq:cluster-spt} in the plane $(h_1/J,h_2/J)$ comprises
the SPT and non-SPT regions identified in Ref.~\cite{cong2019quantum}. The task here is to classify the given ground states as SPT or non-SPT. 

\begin{figure}[!htbp]
\centering
\includegraphics[width=0.95\textwidth]{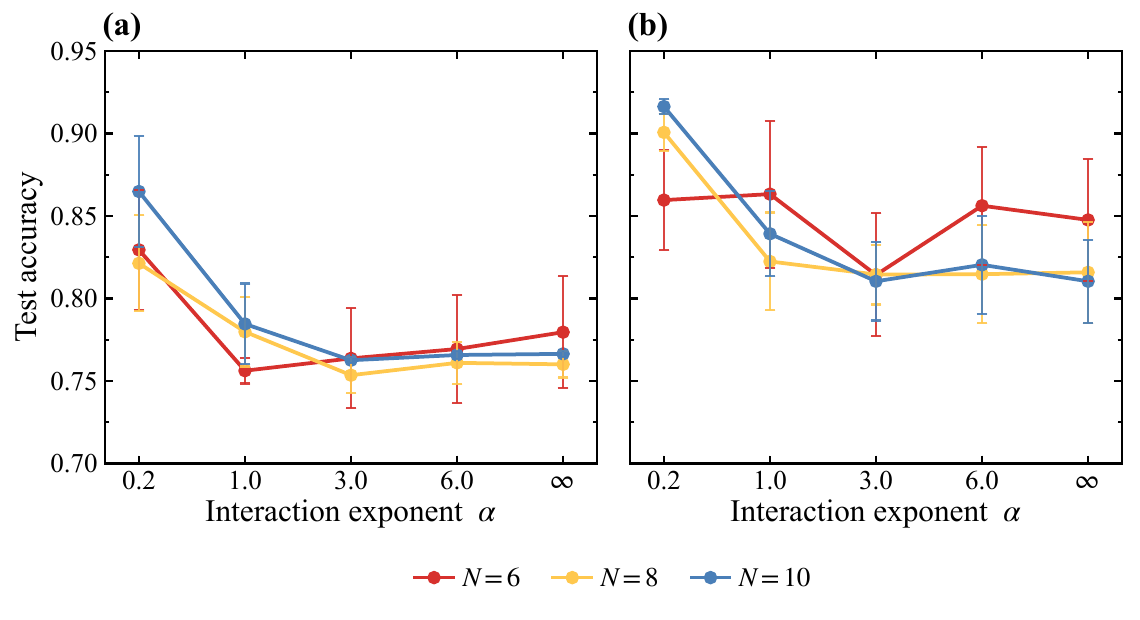}
\caption{Cluster-SPT phase detection accuracy versus the interaction exponent $\alpha$ for $N\!\in\!\{6,8,10\}$ at (a)~$\ell\!=\!1$ and (b)~$\ell\!=\!2$, with $h\!=\!1.0$ and $\tau\!=\!0.4\pi$. Error bars are $95\%$ confidence intervals of the mean over random seeds. }
\label{fig:haldane-phase}
\end{figure}

The circuit architecture  follows
Section~\ref{sec:protocol}: each ground state is loaded directly as the input state,
the ansatz of Eq.~\eqref{eq:ansatz} is applied, and the SPT probability is read
from $q_1$. Training minimizes the cross-entropy loss with labels $y=0$ for non-SPT and $y=1$ for SPT, using AdaGrad to update the circuit rotation angles.
Figure~\ref{fig:haldane-phase} shows the test accuracy as a function of $\alpha$ for $N\!\in\!\{6,8,10\}$ at $\ell\!=\!1$ and $\ell\!=\!2$, revealing how the interaction range changes the results for different system sizes. At $\ell\!=\!2$, the accuracy for $N\!=\!6$ depends only weakly on $\alpha$, but this dependence becomes more significant as $N$ increases. At $N\!=\!10$, the long-range ansatz maintains high accuracy at small $\alpha$, whereas the performance decreases substantially toward the short-range limit for both circuit depths. These size-dependent trends highlight the advantage of using long-range interactions in analog blocks, particularly as systems scale up to the sizes needed for future applications. This advantage is again consistent with the fast scrambling property of long-range interacting dynamics. Further details on the datasets, training procedure, and predicted SPT probabilities are provided in Appendix~\ref{app:spt}.

\subsection{Unsupervised phase detection on an XXZ chain}
\label{sec:xxz}

As an example of unsupervised tasks, following Ref.~\cite{lu2024digital}, we consider a periodic XXZ chain with up to next-nearest-neighbor interactions,
\begin{align}
H=\sum_{i=1}^N\Bigl[
J_3\bigl(\sigma^x_i\sigma^x_{i+1}+\sigma^y_i\sigma^y_{i+1}
+\lambda(\sigma^x_i\sigma^x_{i+2}+\sigma^y_i\sigma^y_{i+2})\bigr)
+J_6\bigl(\sigma^z_i\sigma^z_{i+1}+\lambda^2 \sigma^z_i\sigma^z_{i+2}\bigr)\Bigr],
\label{eq:xxz}
\end{align}
where $J_3$ and $J_6$ are the nearest-neighbor spin-exchange and Ising coupling
strengths, respectively. The relative next-nearest-neighbor strength is $\lambda=(d_{\rm NN}/d_{\rm NNN})^{3}$, with $d_{\rm NN}$ and
$d_{\rm NNN}$ the nearest- and next-nearest-neighbor distances.
We apply the DA-QNN to characterize the ground-state phases of Eq.~\eqref{eq:xxz} in the $(J_3/J_6,\lambda)$ plane, which contains four phases: zAFM, qzAFM, XY-QLRO, and VBS~\cite{mudry2019quantum,lee2023landau}. For each phase, a separate circuit is trained on a single representative ground state by minimizing the loss defined as $L=\frac{1}{N}\sum_{i=1}^{N}p_i(0)$, where $p_i(0)$ is the probability of measuring qubit $i$ in  $|0\rangle$ after the circuit execution. The trained ansatz is then evaluated across a $50\!\times\!50$ grid in parameter space to obtain loss maps. For each random seed $s$, the loss $L_s$ is min--max normalized over the test grid to obtain $\tilde L_s$, and the normalized maps are then averaged over seeds.

\begin{figure}[!htbp]
\centering
\includegraphics[width=\textwidth]{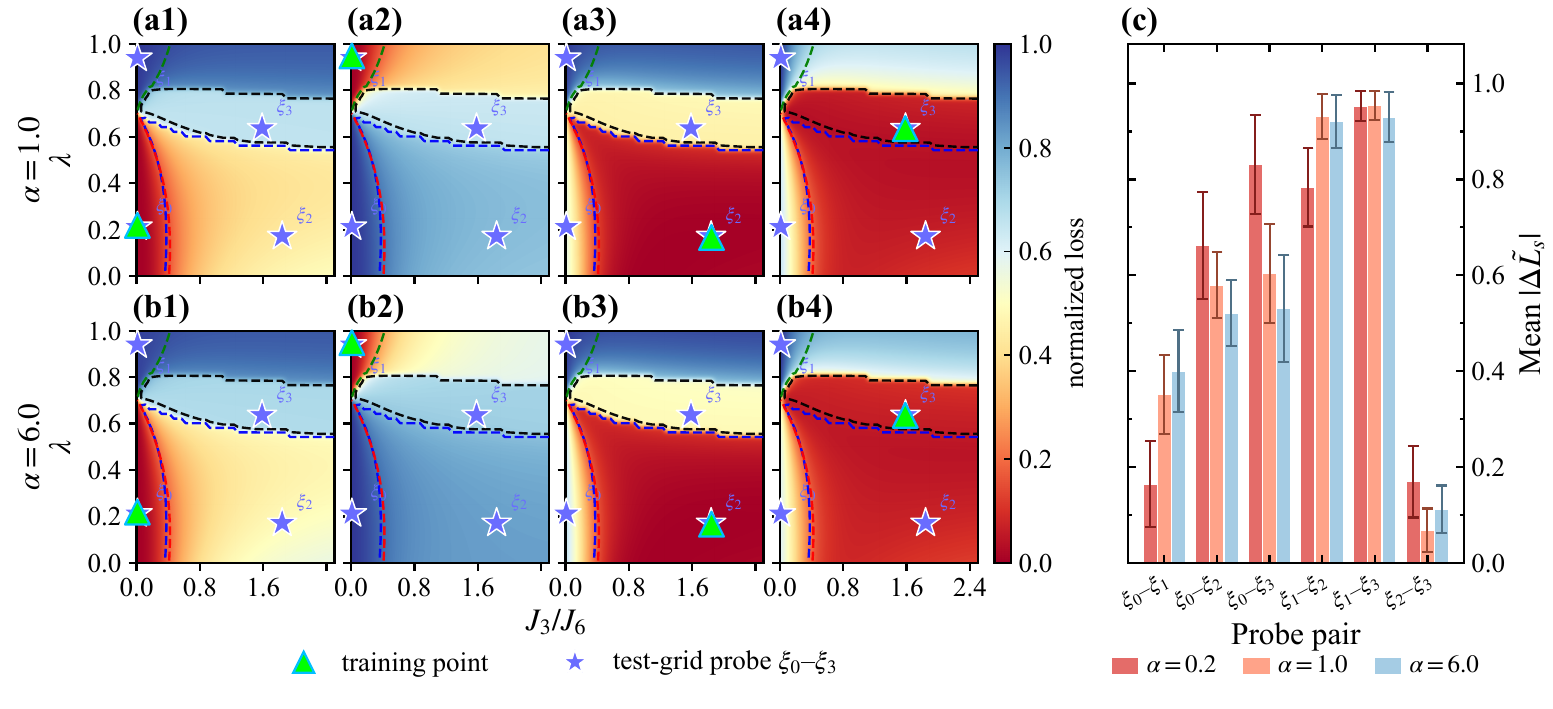}
\caption{Unsupervised XXZ results on the $(J_3/J_6,\lambda)$ plane
($N\!=\!8$, $\ell\!=\!2$, $h\!=\!1.0$, $\tau\!=\!0.4\pi$, averaged over random seeds).
(a1)--(a4) and (b1)--(b4)~show the normalized loss maps for $\alpha\!=\!1.0$ and $\alpha\!=\!6.0$, respectively. In each row, the four columns correspond to circuits trained on ground states from the zAFM, qzAFM, XY-QLRO, and VBS phases. Green triangles mark the training points, while purple stars mark the four probes $\xi_0$, $\xi_1$, $\xi_2$, and $\xi_3$, each selected as the nearest test-grid point to its corresponding training point. Dashed curves indicate phase boundaries obtained from the corresponding order parameters. 
(c)~Mean pairwise absolute differences  in normalized loss over $10$ random seeds, $\operatorname{Mean}|\Delta\tilde L_s|$, for the six pairs among probes $\xi_0$ to $\xi_3$, using circuits trained on the VBS point
($\alpha\!\in\!\{0.2,1.0,6.0\}$). Here, $|\Delta\tilde L_{s}(\xi_i,\xi_j)|=|\tilde L_s(\xi_i)-\tilde L_s(\xi_j)|$. Error bars  are $95\%$ confidence intervals of the mean over random seeds.}
\label{fig:xxz-combined}
\end{figure}

The DA-QNN with long-range interactions shows good performance on this unsupervised quantum phase-boundary-learning task. Here, however, as shown in Figure~\ref{fig:xxz-combined}, it does not consistently outperform its short-range counterpart. Rather, the long-range ansatz provides complementary information, and the two together give a complete and clear picture of the phase diagram. 
Figures~\ref{fig:xxz-combined}(a) and (b) show the learned normalized loss maps for $\alpha\!=\!1.0$ and $\alpha\!=\!6.0$, respectively. The loss maps exhibit phase-dependent structures consistent with the reference order-parameter contours, while their sensitivity to different phases varies with the interaction range. This dependence is further quantified in Figure~\ref{fig:xxz-combined}(c) for the circuit trained on a VBS ground state, through pairwise differences in normalized loss among four probe points $\xi_0$, $\xi_1$, $\xi_2$, and $\xi_3$, each selected near a corresponding training point (zAFM, qzAFM, XY-QLRO, and VBS, respectively). The $\xi_1$--$\xi_3$ pair shows similar loss contrasts across different interaction ranges, whereas the $\xi_0$--$\xi_3$ contrast increases from $0.53$ at $\alpha\!=\!6.0$ to $0.83$ at $\alpha\!=\!0.2$ and the $\xi_0$--$\xi_1$ contrast decreases from $0.40$ to $0.16$. These results demonstrate that long- and short-range interactions provide complementary information for characterizing the four XXZ phases. Further details on the order parameters, training procedure, loss normalization, and probe-pair comparisons are provided in Appendix~\ref{app:xxz}.

\section{Conclusion and outlook}
\label{sec:conclusion}

This work incorporates long-range interacting systems into DA-QNN models and evaluates the performance of the resulting ansatzes across  three types of tasks. Results on binary classification show that incorporating long-range interactions into the analog evolution allows the QNN to achieve good performance with fewer quantum resources. This improvement can be largely attributed to the rapid propagation of information across the analog system facilitated by nonlocal interactions. The resulting advantages are twofold: first, from the perspective of training efficiency, reduced circuit depth alleviates the computational overhead associated with classical optimization; second, in terms of experimental feasibility, shorter execution times generally mitigate the impact of noise in practice. We also note that, both by intuition and by numerical verification, the advantage of long-range ansatzes should be amplified as the system size increases.

Applications to many-body phase detection further demonstrate the value of incorporating long-range interactions into the DA-QNN. With shallow circuits, ansatzes with long-range interacting blocks generally achieve higher accuracy than their short-range counterparts in the cluster-SPT task. In practice, preparing the ground states~\cite{poulin2009preparing,takeuchi2021hardness} already consumes a long time in circuit execution and accordingly exposes the system to accumulated noise. Only if detection is sufficiently fast will it be of practical significance in such a task. In the example of the XXZ chain, there are a few regimes where the long-range ansatzes perform worse than the short-range ones. However, both ansatzes have their own advantageous regimes in this task, and together the complementary results provide a full diagnostic of the phase boundaries.

In general, DA-QNNs do not rely on massive numbers of two-qubit gates, which makes them more experimentally feasible. Incorporating long-range interactions further strengthens the advantages of the analog part by enabling information to propagate more efficiently throughout the system. Beyond the supervised and unsupervised tasks considered here, analog evolutions with long-range interactions should also play an important role in other variational quantum algorithms, such as VQE~\cite{michel2023blueprint} and the quantum approximate optimization algorithm (QAOA)~\cite{tibaldi2025analog}. Although they will not overcome the scaling problem of barren plateaus, they can mitigate this difficulty. Experimentally, we anticipate that DA-QNNs will integrate natively with leading platforms such as neutral atoms, trapped ions, and cavity-QED systems. Algorithmically, long-range interactions are expected to offer advantages in derivative-free protocols, such as quantum reservoir computing~\cite{fujii2017harnessing,xia2022reservoir,lau2024modular} and hybrid quantum simulation~\cite{katz2025hybrid,garciadeandoin2026hamiltonian}. Problem-oriented DA design will be another valuable direction, \textit{e.g.}, identifying suitable analog Hamiltonians based on the topological or entanglement structure required by a problem. Moreover, hardware-native nonlocal interactions are not limited to power-law forms: movable optical tweezers in neutral-atom arrays~\cite{losl2026variational} and multi-beam combinations in trapped ions~\cite{richerme2025multimode,zhang2026residual,huang2026large} can provide different forms of nonlocal interactions. These options will further enrich digital-analog protocols.

\section*{Contribution}
Z.G. developed the QNN code and performed the MNIST classification; D.L. carried out the many-body phase detection; and Z.Z. supervised the project and provided computing resources.

\ack{The authors thank Jingze Zhuang for helpful discussions. This work is supported by the Shanghai Pujiang Program under Grant No.~25PJA110.}

\data{The data and code that support the findings of this study are openly available at
\url{https://github.com/HUSH-cloud/DA-QNN-long-range-interactions}.
}

\bibliographystyle{iopart-num}
\bibliography{refs}
\newpage

\appendix
\numberwithin{equation}{section}
\makeatletter
\def\appendixfmt@section{Appendix~\thesection.\quad}
\def\appendixfmt@subsection{\thesubsection\quad}
\def\appendixfmt@subsubsection{\thesubsubsection\quad}
\renewcommand{\@seccntformat}[1]{\csname appendixfmt@#1\endcsname}
\makeatother

\section{Task difficulty of the digit pairs}
\label{app:task-diff}

\begin{figure}[!htbp]
\centering
\includegraphics[width=0.6\textwidth]{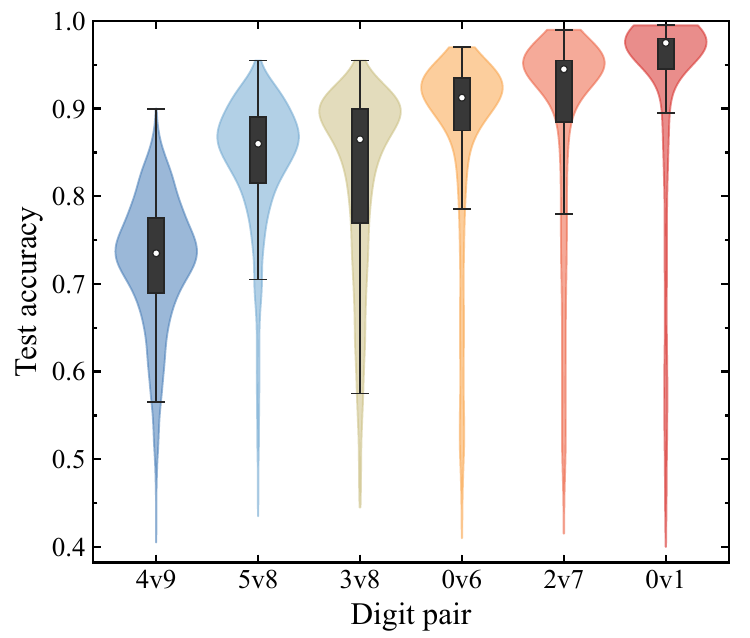}
\caption{Test accuracy distribution per MNIST digit pair, aggregated over
$(\ell,N,\alpha)$ and random seeds at $h\!=\!1.0$. The shaded outline is a kernel density estimate over the
runs. Inside it the bar spans the interquartile range (IQR), the whiskers reach
the extreme runs within $1.5 \times $IQR of the quartiles, and the dot marks
the median.}
\label{fig:task-diff}
\end{figure}

Choosing an appropriate task is important for isolating the role of the interaction range. Here we scan the performance of DA-QNNs over a wide range of parameters for six representative  MNIST digit
pairs, and the results are summarized in Figure~\ref{fig:task-diff}.
Each distribution pools the test accuracy at
$\tau\!=\!0.4\pi$ over $\ell\!\in\!\{1,2,4\}$,
$N\!\in\!\{3,\dots,10\}$ and $\alpha\!\in\!\{0.2,1.0,3.0,6.0,\infty\}$ at a 
transverse field of $h\!=\!1.0$, with 20 random seeds each (in total $2400$ runs per pair).
The median accuracies for these pairs are $0.74$ for 4v9, $0.86$ for 5v8, $0.87$ for 3v8, $0.91$ for 0v6, $0.95$ for 2v7, and $0.98$ for 0v1.
Among all these tasks, the 3v8 pair is of moderate difficulty and well suited to the aim of this work: it is neither so easy that all ansatzes saturate nor so hard that all of them fail to learn, thus avoiding both ceiling and floor effects.

\section{Physical parameters}
\label{app:field-range}

The transverse field $h$ and the power-law exponent $\alpha$ act on the analog
block together, with $\alpha$ setting how fast and how far correlations spread and $h$, together with $\alpha$, setting how quickly the system thermalizes. This appendix discusses how $h$ and $\alpha$ control thermalization in the analog evolution, and how they jointly determine the performance of the DA-QNN.

\subsection{Thermalization of the analog block}
\label{app:therm}

To evaluate the thermalization timescale of the analog evolution, we present the
magnetization dynamics of the LR-TFIM Hamiltonian. The system is
initialized in a fully polarized product state
$\ket{\uparrow}^{\otimes N}$ (the computational basis state
$\ket{0}^{\otimes N}$), evolved under the analog Hamiltonian
$H$ of Eq.~\eqref{eq:H} with $N\!=\!8$, and measured with the normalized magnetization $M_z(\tau)=\tfrac{1}{N}\sum_i\langle
\sigma^z_i(\tau)\rangle$ over time. We report its accumulated magnetization
$\overline{M}_z(\tau)=\tfrac{1}{\tau}\int_0^{\tau}M_z(\tau')\,d\tau'$ as an indicator of thermalization. Here $\overline{M}_z\!\to\!0$ marks 
thermalization toward the disordered state. The scan runs over
$\tau\!\in\![0,6\pi]$ on a $4000$-point time grid. 

\begin{figure}[h]
\centering
\includegraphics[width=0.6\textwidth]{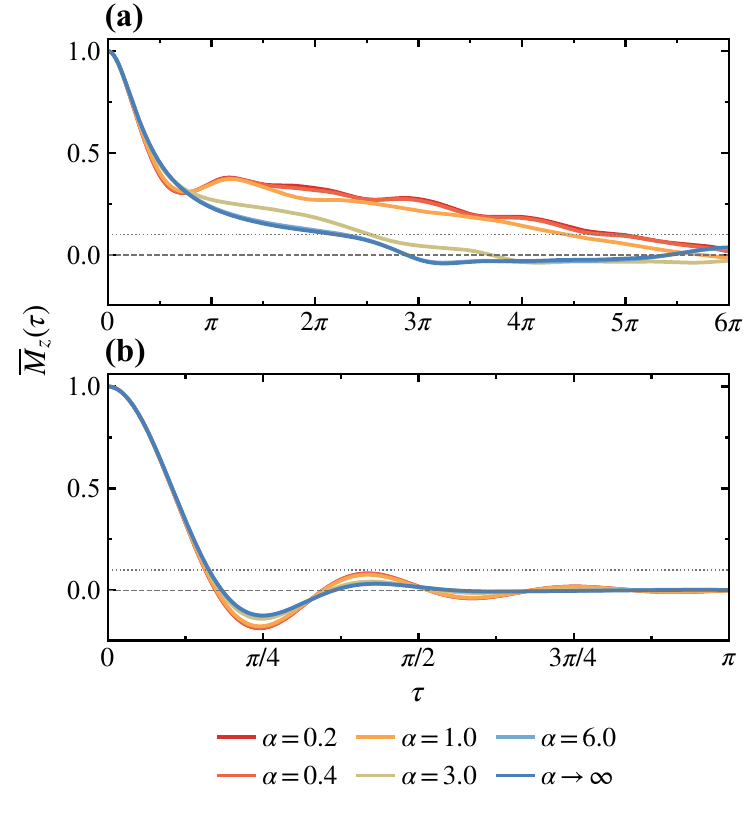}
\caption{Time-accumulated magnetization
$\overline{M}_z(\tau)=\tfrac{1}{\tau}\int_0^{\tau}\!M_z(\tau')\,d\tau'$ for six
$\alpha$'s at $N\!=\!8$, initialized from $\ket{\uparrow}^{\otimes N}$.
(a)~$h\!=\!1.0$ over $\tau\!\in\![0,6\pi]$. (b)~$h\!=\!3.0$ over
$\tau\!\in\![0,\pi]$. Dotted line: $\overline{M}_z\!=\!0.1$, dashed line: $\overline{M}_z\!=\!0$.}
\label{fig:therm}
\end{figure}

As shown in Figure~\ref{fig:therm}, the long-range interacting models show prethermalization at small $h$, which is consistent with the results in Refs.~\cite{gong2013prethermalization,mori2019prethermalization}. At $h\!=\!1.0$, the long-range systems ($\alpha\!\lesssim\!1.0$)
retain a sufficiently large accumulated magnetization, forming a prethermal plateau with
$\overline{M}_z$ staying well above zero up to $\tau\!\approx\!4\pi$ with a
slow secondary revival near $\tau\!\sim\!\pi$, whereas the
short-range limit ($\alpha\!\to\!\infty$) relaxes much more quickly
toward $\overline{M}_z\!\to\!0$. This prethermal plateau offers an advantage in learning tasks, in that the learning power of the DA-QNN can be robust against the choice of $\tau$ with long-range interactions. Besides prethermalization, long-distance
correlations can still be rapidly built~\cite{lerose2020origin}.
At a large transverse field of $h\!=\!3.0$, evolution with all $\alpha$ thermalizes rapidly, as $\overline{M}_z$ drops to
zero within $\tau\!\lesssim\!\pi$.

\subsection{Binary classification accuracy across \texorpdfstring{$(h,\alpha)$}{(h, alpha)} values}
\label{app:h-alpha}

\begin{figure}[!htbp]
\centering
\includegraphics[width=0.65\textwidth]{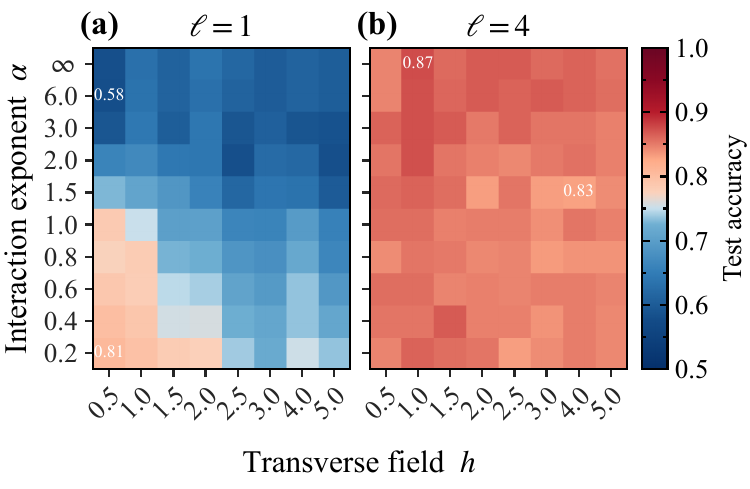}
\caption{Binary classification accuracy of the digit pair 3v8 across $(h,\alpha)$ values for
(a)~$\ell\!=\!1$ and (b)~$\ell\!=\!4$ at $N\!=\!8$ and $\tau\!=\!0.4\pi$.
Results are averaged over random seeds. The best and worst cells of each panel are annotated.}
\label{fig:h-alpha-heatmap}
\end{figure}

Here we present a parameter scan across the $(h,\alpha)$ values, complementing the parameter scan over $(\tau,\ell)$ in Section~\ref{sec:fixed-tl} of the main text. With $\tau=0.4\pi$ fixed and two depths $\ell=1$ and $4$, the MNIST task 3v8 is classified with 
$h\in[0.5,\allowbreak 5.0]$ and
$\alpha\in\{0.2,\allowbreak 0.4,\allowbreak 0.6,\allowbreak 0.8,\allowbreak
1.0,\allowbreak 1.5,\allowbreak 2.0,\allowbreak 3.0,\allowbreak 6.0,\allowbreak\infty\}$, and the results are summarized in Figure~\ref{fig:h-alpha-heatmap}.
For $\ell\!=\!1$, the regime of small $\alpha$ and $h$ yields high accuracy. For short-range interactions with $\alpha>1$, a single entangling layer is insufficient for information to propagate across the whole system, which largely limits expressibility. A transverse field that is too strong can likewise hinder information propagation, since it tends to align all qubits along its direction.
For $\ell\!=\!4$, the entangling layers are sufficient to allow information to propagate fully across the system, so that performance is good over the entire parameter space. Performance at very large $h$ is slightly worse than in the small-$h$ regime, since the regime deep in the  disordered phase may not be a favorable setting for learning tasks.

\section{Minimal resource estimate }
\label{app:resource-curves}

\begin{figure}[!htbp]
\centering
\includegraphics[width=0.65\textwidth]{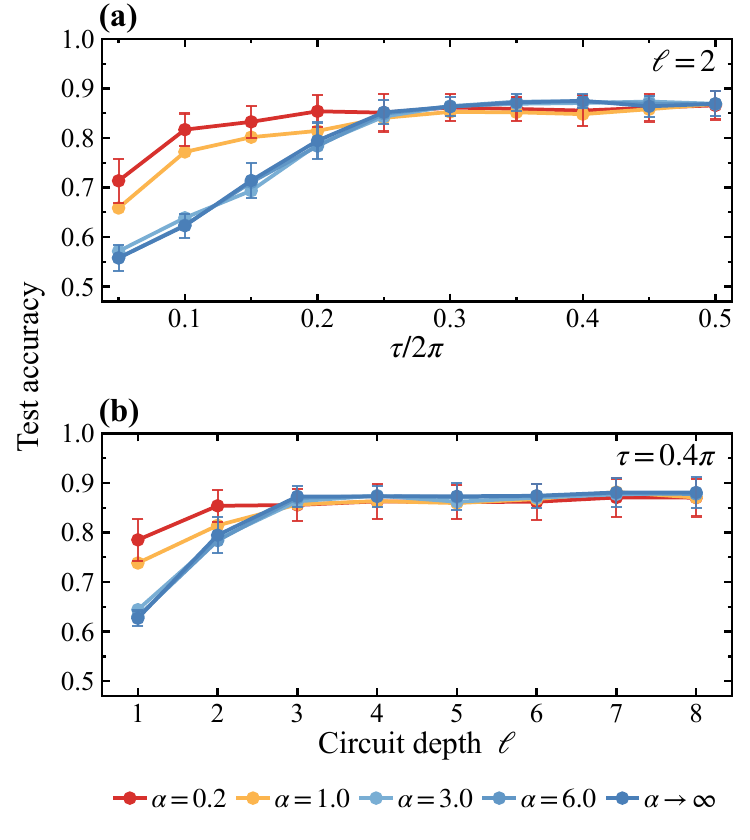}
\caption{Test accuracy versus (a)~the layerwise evolution time $\tau$ at $\ell\!=\!2$
and (b)~the circuit depth $\ell$ at $\tau\!=\!0.4\pi$, with $N\!=\!8$ and $h\!=\!1.0$.
Error bars are $95\%$ confidence intervals of the mean over random seeds and shown only for $\alpha\!=\!0.2$ and $\alpha\!\to\!\infty$ for clarity.}
\label{fig:resource-curves}
\end{figure}

This appendix provides detailed information on Section~\ref{sec:depth-time} of the main text, the main message of which is  summarized in
Figure~\ref{fig:resource}. Figures~\ref{fig:resource-curves}(a) and (b) respectively plot the  test accuracy for two settings: (a) fixed $\ell=2$, with $\tau$ scanned from $0.1\pi$ to $\pi$; (b) fixed $\tau=0.4\pi$, with $\ell$ scanned from 1 to 8.
In each plot, we show the five $\alpha$ values at
$N\!=\!8$ and $h\!=\!1.0$. Both panels show the same qualitative behavior: DA-QNNs with long-range interactions perform better than those with short-range interactions at  low cost in quantum resources, which can be either the running time of the analog blocks or the number of QNN layers. However, when the quantum resources are sufficient, the differences between ansatzes with different $\alpha$ are within the error bars.

\section{Results at greater circuit depth}
\label{app:depth}

Here we present results for deeper circuits than those  in Sections~\ref{sec:alphaN} and \ref{sec:encoding} by repeating the size  and encoding comparisons at $\ell=4$. Intuitively, for deeper circuits, the differences among ansatzes with different interaction ranges should smear out, but this does not weaken our conclusions drawn for shallow circuits.

\subsection{Size-dependent comparison}
\label{app:depth-size}

\begin{figure}[!htbp]
\centering
\includegraphics[width=0.65\textwidth]{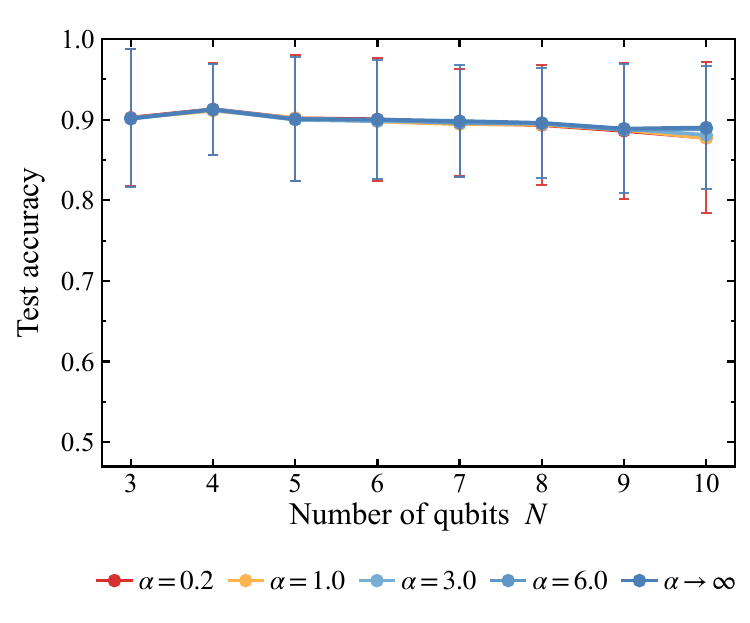}
\caption{Binary classification accuracy with various system sizes $N$ at $\ell\!=\!4$ and $h\!=\!1.0$, averaged over six digit pairs. Error bars are one standard deviation across the six pairs and random seeds and shown only for $\alpha\!=\!0.2$ and $\alpha\!\to\!\infty$ for clarity.}
\label{fig:alphaN-ell4}
\end{figure}

Here we present the size-dependent comparison at $\ell=4$, complementing the results  in Section~\ref{sec:alphaN} of the main text. With $\tau=0.4\pi$ and $h=1.0$ fixed, the accuracy is averaged over the six digit pairs for $N\in\{3,\dots,10\}$, and the results are summarized in Figure~\ref{fig:alphaN-ell4}. For all  tested system sizes, four sequential analog blocks are enough for information to propagate across the whole system for every $\alpha$, so that the accuracy shows no dependence on the interaction range. Combined with Figure~\ref{fig:alphaN-acc}, these results show that the system sizes at which the performance is independent of $\alpha$ grow with the circuit depth, from $N\le 5$ at $\ell=1$ and $N\le 7$ at $\ell=2$ to at least $N=10$ at $\ell=4$. Following this trend, the advantage of the long-range ansatz is expected to persist at larger $N$. The slight decrease of accuracy with $N$ is shared by all $\alpha$ and has the same origin as discussed in Section~\ref{sec:alphaN}. These results are consistent with the saturation over $\ell$ in Figure~\ref{fig:tl-heatmap} and the uniform performance at $\ell=4$ in Appendix~\ref{app:h-alpha}.

\subsection{Encoding strategies}
\label{app:depth-encoding}

\begin{figure}[!htbp]
\centering
\includegraphics[width=0.65\textwidth]{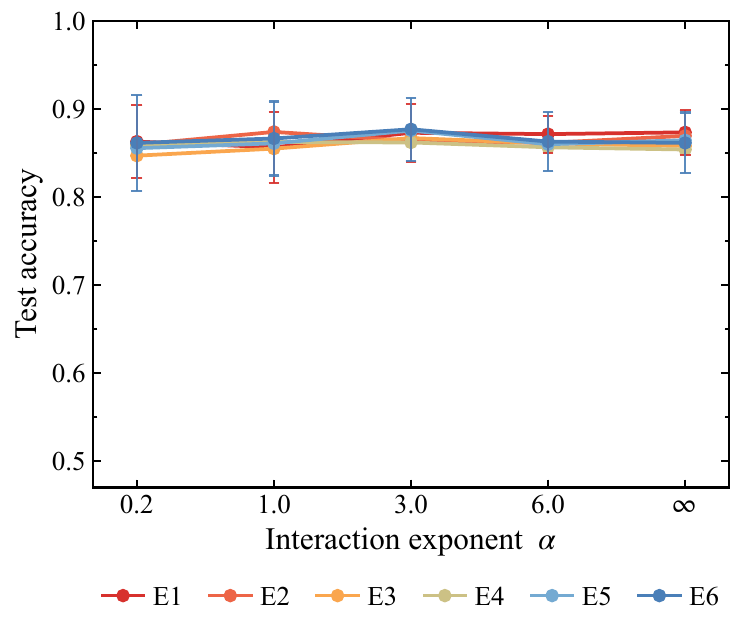}
\caption{Effect of the encoding strategy on binary classification accuracy across $\alpha$ for the pair 3v8 at $N\!=\!8$, $h\!=\!1.0$, and $\ell\!=\!4$. Error bars are $95\%$ confidence intervals of the mean over random seeds and shown only for E1 and E6 for clarity.}
\label{fig:encoding-ell4}
\end{figure}

Here we present the comparison of encoding strategies at $\ell=4$, complementing the results  in Section~\ref{sec:encoding} of the main text. With $N=8$, $h=1.0$, and $\tau=0.4\pi$ fixed, the six encoding strategies defined in Figure~\ref{fig:encoding-ablation} are compared, and the results are summarized in Figure~\ref{fig:encoding-ell4}. Unlike the case of $\ell=1$, where the short-range ansatzes are sensitive to the encoding, all six encoding strategies yield similar accuracies for every $\alpha$ at $\ell=4$. This supports the conclusion in Section~\ref{sec:encoding} that the choice of encoding matters only for shallow circuits.

\section{Diagnostics for the quantum-state tasks}
\label{app:phase-tasks}

This part provides supplemental information for the cluster-SPT and XXZ tasks in 
Section~\ref{sec:phase-detection} of the main text. 

\subsection{Supervised phase classification on a cluster-SPT chain}
\label{app:spt}

Figure~\ref{fig:haldane-phase} uses a balanced training set of $50$ ground
states ($25$ for SPT and $25$ for non-SPT). The ground states for $N\!\in\!\{6,8,10\}$ are labeled using fitted thermodynamic-limit phase boundaries based on Ref.~\cite{cong2019quantum}. A state is labeled SPT when
$h_2/J$ lies between the two fits by more than a margin of $0.01$, and non-SPT
when it lies beyond either of them by more than the same margin. The $25$ non-SPT training states comprise $5$ antiferromagnetic and $20$ paramagnetic states. The circuits are trained for $500$ iterations using AdaGrad to minimize the cross-entropy loss. The test set contains $265$ non-SPT and $176$ SPT ground states at parameter points derived from a $21\!\times\!21$ grid. Test points within the margin are shifted outside it before labeling. Classification uses a threshold of $0.5$ on the predicted SPT probability.

\begin{figure}[!htbp]
\centering
\includegraphics[width=0.65\textwidth]{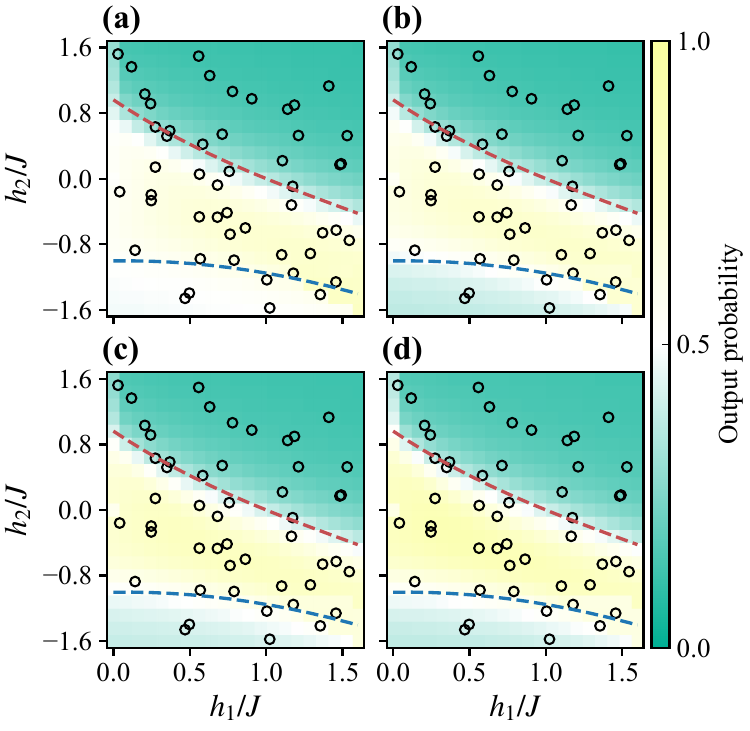}
\caption{Predicted SPT probabilities of the cluster-SPT task on $(h_1/J,h_2/J)$ from the DA-QNN with $N\!=\!8$ and $\alpha\!=\!0.2$, averaged over random seeds. Circles mark the $50$ random training states and the
dashed curves are the two phase boundaries used for labeling. The region between the dashed curves is the SPT phase, while the upper and lower regions correspond to the paramagnetic and antiferromagnetic phases, respectively.
(a)~$\ell\!=\!1$, (b)~$\ell\!=\!2$, (c)~$\ell\!=\!4$, (d)~$\ell\!=\!8$.}
\label{fig:haldane-prob}
\end{figure}

Figure~\ref{fig:haldane-prob} shows the predicted SPT probabilities
on the $(h_1/J,h_2/J)$ plane obtained from the DA-QNN with $\alpha\!=\!0.2$ at different $\ell$.
At $\ell\!=\!1$, the learned boundary is blurred. As the number of QNN layers increases,
the learned  boundaries sharpen. Eventually, at $\ell\!=\!8$, the phase detection accuracy rises to $0.93$, and the learned phase boundary lies close to the reference boundaries. The classifier therefore learns a spatially
organized phase partition rather than isolated training points.

\subsection{Unsupervised phase detection on an XXZ chain}
\label{app:xxz}

\begin{figure}[!htbp]
\centering
\includegraphics[width=0.6\textwidth]{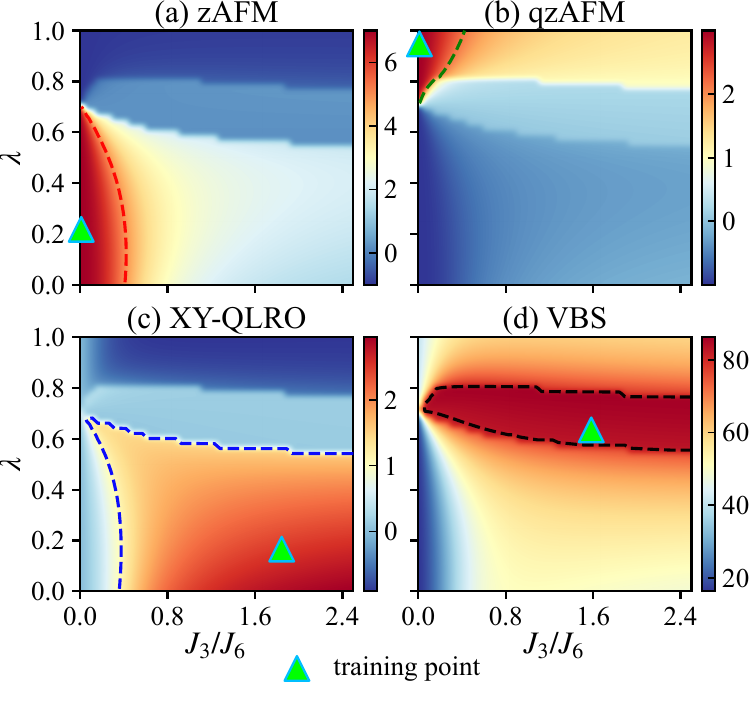}
\caption{XXZ order-parameter fields on the $(J_3/J_6,\lambda)$ plane at
$N\!=\!8$: (a)~zAFM, (b)~qzAFM, (c)~XY-QLRO, (d)~VBS. Each color map shows the
corresponding order parameter. The dashed curve is its quantile iso-contour,
used as  an approximate phase boundary in Figure~\ref{fig:xxz-combined}. In each
panel, the green triangle marks the training point of that phase.}
\label{fig:xxz-order}
\end{figure}

\begin{figure}[!htbp]
\centering
\includegraphics[width=0.65\textwidth]{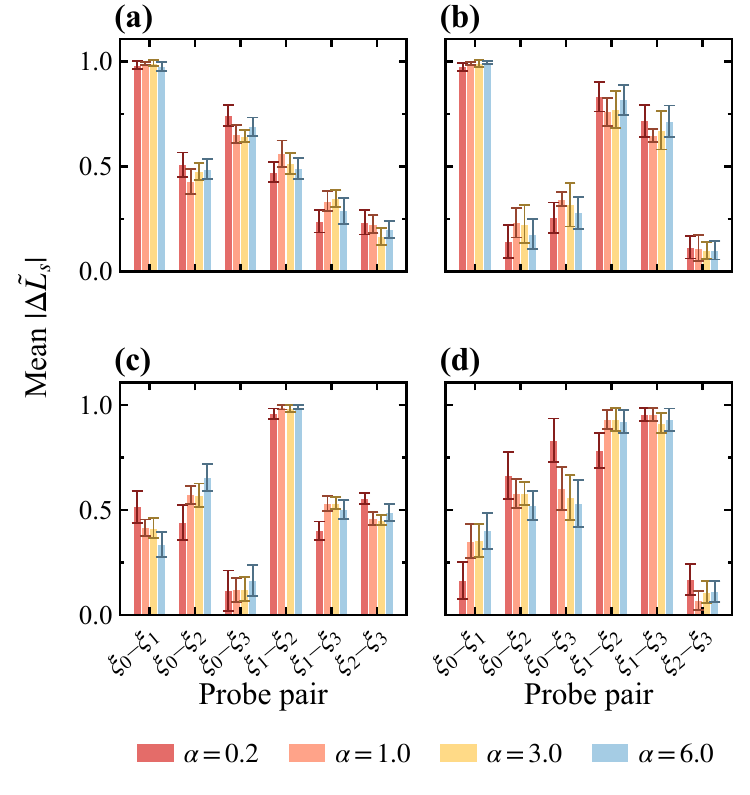}
\caption{Pairwise absolute differences of the normalized loss $\tilde L_s$ among
probes $\xi_0$ to $\xi_3$ ($N\!=\!8$, $\ell\!=\!2$, $h\!=\!1.0$, $\tau\!=\!0.4\pi$, averaged over random seeds).
(a)--(d)~zAFM, qzAFM, XY-QLRO, and VBS training points. Error bars are $95\%$ confidence intervals of the mean over random seeds.}
\label{fig:xxz-pairdiff}
\end{figure}

For the XXZ task, Figure~\ref{fig:xxz-order} shows the order-parameter fields
used to draw the dashed boundaries in Figure~\ref{fig:xxz-combined}. The four phases are a $z$-axis antiferromagnet with N\'eel order (zAFM), a longitudinal phase with period-four order (qzAFM), an XY phase with quasi-long-range transverse order (XY-QLRO), and a valence-bond solid (VBS).  The first three order-parameter fields are constructed from two-point correlators referenced to the first qubit,
\begin{equation}
\mathcal{O}_\mu(k)=\mathrm{Re}\sum_{r=1}^{N-1} e^{ikr}\,
\bigl\langle \sigma^{\mu}_{1}\,\sigma^{\mu}_{1+r}\bigr\rangle ,
\label{eq:xxz-op}
\end{equation}
with $(\mu,k)=(z,\pi)$ for zAFM, $(z,\pi/2)$ for qzAFM, and $(x,\pi)$ for
XY-QLRO. The fourth is built from the bond-difference operator
$D_j=\bm{\sigma}_j\!\cdot\!\bm{\sigma}_{j+1}
-\bm{\sigma}_{j-1}\!\cdot\!\bm{\sigma}_{j}$,
\begin{equation}
\mathcal{O}_{\rm VBS}=\mathrm{Re}\sum_{r=1}^{N}(-1)^{r}\,
\bigl\langle D_{1}D_{1+r}\bigr\rangle ,
\label{eq:xxz-vbs}
\end{equation}
with site indices taken modulo $N$ throughout. Since the XY phase has no local order
parameter in the thermodynamic limit, we use the staggered
transverse correlator $\mathcal{O}_x(\pi)$ from Eq.~\eqref{eq:xxz-op} as a
finite-size proxy for its
quasi-long-range order. Each dashed contour is a single iso-level of one of
these fields, placed at the $f$-quantile of that field with $f=0.50$ (XY-QLRO), $0.82$ (VBS), $0.91$ (zAFM), and $0.97$ (qzAFM).
These iso-contours serve as finite-size references for the corresponding phase boundaries.
Each boundary is drawn in the color of its phase: red (zAFM), green
(qzAFM), blue (XY-QLRO), and black (VBS), both here and in
Figure~\ref{fig:xxz-combined}. Green triangles mark the four training points at $(0.01,0.2184)$, $(0.01,0.9479)$, $(1.8447,0.1663)$, and $(1.5826,0.6353)$ for the zAFM, qzAFM, XY-QLRO, and VBS phases, respectively. The corresponding ground states are used directly as circuit inputs without adiabatic preparation. Each circuit is trained using AdaGrad for $50$ iterations.

For each random seed $s$, the loss $L_s$ is min--max normalized over the $50\!\times\!50$ test grid as
\begin{equation}
\tilde L_s=\frac{L_s-\min_{\rm grid}L_s}{\max_{\rm grid}L_s-\min_{\rm grid}L_s}.
\end{equation}
The normalized loss maps are averaged over $S=10$ random seeds. The four probes $\xi_0$, $\xi_1$, $\xi_2$, and $\xi_3$ are the test-grid points nearest to the corresponding training points. For each probe pair $(\xi_i,\xi_j)$, the mean absolute loss difference shown in Figures~\ref{fig:xxz-pairdiff} and \ref{fig:xxz-combined}(c) is
\begin{equation}
\operatorname{Mean} \left|\Delta\tilde L_s\right|
=\frac{1}{S}\sum_{s=1}^{S}
\left|\tilde L_s(\xi_i)-\tilde L_s(\xi_j)\right|,
\end{equation}
where $\Delta\tilde L_s=\tilde L_s(\xi_i)-\tilde L_s(\xi_j)$. Error bars indicate $95\%$ $t$-based confidence intervals for the mean absolute loss differences across seeds.

Figure~\ref{fig:xxz-pairdiff} compares the mean absolute loss differences for the four training points at
$\ell\!=\!2$ and $\alpha\!\in\!\{0.2,1.0,3.0,6.0\}$. The dependence on the interaction range varies with both the training point and the probe pair. For the zAFM and qzAFM training points, the loss differences vary only modestly with $\alpha$ for most probe pairs, whereas XY-QLRO and VBS exhibit more pronounced pair-dependent variations. For XY-QLRO in Figure~\ref{fig:xxz-pairdiff}(c), the $\xi_0$--$\xi_1$ mean loss difference is larger at $\alpha=0.2$ than at $\alpha=6.0$, whereas the opposite holds for $\xi_0$--$\xi_2$. Together with the VBS results in Figure~\ref{fig:xxz-combined}(c), these results highlight the roles of ansatzes with long- and short-range interactions in XXZ phase characterization.

\end{document}